\documentstyle[12pt]{article}
\vspace{1.8cm} \oddsidemargin 0pt \evensidemargin 0pt  \topmargin
-1.5cm \footheight 26pt \footskip 10mm

\begin{document}
\baselineskip 20pt
\begin{center}
\baselineskip=24pt {\LARGE Special Theory for Superluminal
Particle}

\vspace{1cm} \centerline{Xiang-Yao Wu$^{a}$
\footnote{E-mail:wuxy2066@163.com}, Bo-Jun Zhang$^{a}$, Xiao-Jing
Liu$^{a}$, Nuo Ba$^{a}$ and Yan Wang$^{a}$ }

\vskip 10pt \noindent{\footnotesize a. \textit{Institute of
Physics, Jilin Normal University, Siping 136000, China}}

\end{center}
\date{}
\renewcommand{\thesection}{Sec. \Roman{section}} \topmargin 10pt
\renewcommand{\thesubsection}{ \arabic{subsection}} \topmargin 10pt
{\vskip 5mm
\begin {minipage}{140mm}
\centerline {\bf Abstract} \vskip 8pt
\par

\indent\\

\hspace{0.3in} The OPERA collaboration reported evidence for
muonic neutrinos travelling faster than light in vacuum. In this
paper, an extended relativity theory is proposed. We think all
particles can be divided into three kinds: The first kind of
particle is its velocity in the range of $0\leq v < c$, e.g.
electron, atom, molecule and so on ($c$ is light velocity, i.e.,
the limit velocity of the first kind of particle). The second kind
of particle is its velocity in the range of $0\leq v < c_{m1}$,
e.g. photon ($c_{m1}$ is the limit velocity of the second kind of
particle). The third kind of particle is its velocity in the range
of $c\leq v < c_{m2}$, e.g. tachyon, and muonic neutrinos
($c_{m2}$ is the limit velocity of the third kind of particle).
The first kind of particle is described by the special relativity.
With the extended relativity theory, we can describe the second
and third kinds particles, and can analysis the OPERA experiment
results and calculate the muonic neutrinos mass.

\vskip 5pt
PACS numbers: 03.30.+p; 42. 90. +m; 03.65.Pm \\

Keywords: Special relativity; Superluminal particle; Quantum
theory
\end {minipage}

\newpage
\section * {1. Introduction }

\hspace{0.3in}Recently the OPERA collaboration reported evidence
for muonic neutrinos slightly faster than light in vacuum [1]. The
CERN Neutrino beam to Gran Sasso (CNGS) consists of $\nu_{\mu}$,
with small contaminations of $\bar{\nu_{\mu}}$ ($2.1\%$) and of
$\nu_{e}$ or $\bar{\nu_{e}}$ (together less than 1\%). At the
average neutrino energy $17$ GeV, the relative difference of the
velocity of the muon neutrinos $v$ with respect to light quoted by
OPERA is: $(v-c)/c=(2.48\pm0.28(stat)\pm0.30(sys))\times 10^{-5}$.
The velocity measurement of the muon neutrinos were also reported
for muon neutrino beams produced at Fermilab. Dealing with
energies peaked at $3$ GeV, the MINOS Collaboration [2] found in
2007 that $(v-c)/c=(5.1\pm2.9)\times 10^{-5}$. The measurement
results above are seemingly in conflict with special relativity in
$4$ dimensions, a number of possible explanations for Lorentz
violation exist in the literature [3-12]. The neutrino velocity is
higher than the velocity of light, but we think there is a limit
velocity in the nature. In the paper, we propose a full relativity
theory, which is based on the two postulates:\\
1. All particles can be divided into three kinds: The first kind
is its velocity in the range of $0\leq v < c$, e.g. electron,
atom, molecule and so on, the light velocity $c$ is their limit
velocity. The second kind is its velocity in the range of $0\leq v
< c_{m1}$($c< c_{m1}$), e.g. photon, the velocity $c_{m1}$ is its
limit velocity. The third kind is its velocity in the range of
$c\leq v < c_{m2}$, e.g. muonic neutrinos and tachyon. The
velocity
$c_{m2}$ is their limit velocity.\\
2. The first kind of particle is described by Einstein's special
relativity.

In the paper, we shall study the second and third kinds of
particles by the extended relativity theory. With the extended
relativity theory, we give new results about photon, and calculate
the limit velocity $c_{m1}$ and $c_{m2}$. Otherwise, we analysis
the OPERA experiment data and calculate the muonic neutrinos mass.

\section * {2. The space-time transformation and mass-energy relation
for the second kind of particle ($0\leq \upsilon < c_{m1}$)}

For the first kind of particle, its velocity is in the range of
$0\leq v< c$, e.g. electron, atom, molecule and so on, and they
can be described by the special relativity. In 1905, Einstein gave
the space-time transformation and mass-energy relation which are
based on his two postulates, i.e., the invariant principle of
light velocity and the relativity principle. The space-time
transformation is
\begin{eqnarray}
x&=&\frac{x^{'}+u
t^{'}}{\sqrt{1-\frac{u^{2}}{c^{2}}}}\nonumber\\
 y&=&y^{'}\nonumber\\
z&=&z^{'}\nonumber\\
t&=&\frac{t^{'}+\frac{u}{c^{2}}x^{'}}{\sqrt{1-\frac{u^{2}}{c^{2}}}},
\end{eqnarray}
where $x$, $y$, $z$, $t$ are space-time coordinates in $\sum$
frame, $x^{'}$, $y^{'}$, $z^{'}$, $t^{'}$ are space-time
coordinates in $\sum^{'}$ frame, $c$ is the speed of light, and
$u$ is the relative velocity between $\sum$ and $\sum^{'}$ frame,
which move along $x$ and $x^{'}$ axes.
The velocity transformation is\\
\begin{eqnarray}
u_{x}&=&\frac{u_{x}^{'}+u
}{1+\frac{u u_{x}^{'}}{c^{2}}}\nonumber\\
u_{y}&=&\frac{u_{y}^{'}\sqrt{1-\frac{u^{2}}{c^{2}}}}
{1+\frac{u u_{x}^{'}}{c^{2}}}\nonumber\\
u_{z}&=&\frac{u_{z}^{'}\sqrt{1-\frac{u^{2}}{c^{2}}}} {1+\frac{u
u_{x}^{'}}{c^{2}}},
\end{eqnarray}
where $u_{x}$, $u_{y}$ and $u_{z}$ are a particle velocity
projection in $\sum$ frame, $u_{x}^{'}$, $u_{y}^{'}$ and
$u_{z}^{'}$ are the particle velocity projection in $\sum^{'}$
frame.

The mass, momentum and energy of a particle of rest mass $m_{0}$
and velocity $\vec{v}$ are:
\begin{equation}
m=\frac{m_{0}}{\sqrt{1-\frac{v^{2}}{c^{2}}}},
\end{equation}
\begin{equation}
\vec{p}=\frac{m_{0}\vec{v}}{\sqrt{1-\frac{v^{2}}{c^{2}}}},
\end{equation}
and
\begin{equation}
E=mc^{2}=\frac{m_{0}}{\sqrt{1-\frac{v^{2}}{c^{2}}}}c^{2},
\end{equation}
and they satisfy the dispersion relation
\begin{equation}
E^{2}=m_{0}^{2}c^{4}+p^{2}c^{2}.
\end{equation}
For the second kind of particle, its velocity is in the range of
$0\leq \upsilon < c_{m1}$ $(c<c_{m1})$, e.g. photon ($c_{m1}$ is
the limit velocity of the second kind of particle). Recently, a
series of experiments revealed that electromagnetic wave was able
to travel at a group velocity faster than $c$. These phenomena
have been observed in dispersive media [13, 14], in electronic
circuits [15], and in evanescent wave cases [16, 17]. It is about
40 years before, that O.M.P. Bilaniuk, V.K. Deshpande and E.S.G.
Sudarshan have studied the space-time relation for superluminal
reference frames within the framework of special relativity [18,
19]. They assumed that the space-time and velocity transformation
of special relativity are suitable for superluminal reference
frames. They obtained the new results that the proper length
$L_{0}$ and proper time $T_{0}$ must be imaginary so that the
measured quantities, such as length $L$ and time $T$, are real. We
think there is superluminal photon, but its velocity can not be
infinity. So, we think there is a limit velocity $c_{m1}$ for the
superluminal photon and give two postulates for the second kind of
particle (photon) as follows:

1. The Principle of Relativity: All laws of nature are the same in
all inertial reference frames.

2. The Universal of Limit Velocity: There is a limit velocity
$c_{m1}$, and the $c_{m1}$ is invariant in all inertial reference
frames.

From the two postulates, we can obtain the space-time
transformation and velocity transformation for the second kind of
particle ($0\leq \upsilon <c_{m1}$). When we replace $c$ with
$c_{m}$, we can obtain the new space-time transformation from the
Lorentz transformation, they are
\begin{eqnarray}
x&=&\frac{x^{'}+u
t^{'}}{\sqrt{1-\frac{u^{2}}{c_{m1}^{2}}}}\nonumber\\
 y&=&y^{'}\nonumber\\
z&=&z^{'}\nonumber\\
t&=&\frac{t^{'}+\frac{u}{c_{m1}^{2}}x^{'}}{\sqrt{1-\frac{u^{2}}{c_{m1}^{2}}}},
\end{eqnarray}
where $x$, $y$, $z$, $t$ are space-time coordinates in $\sum$
frame, $x^{'}$, $y^{'}$, $z^{'}$, $t^{'}$ are space-time
coordinates in $\sum^{'}$ frame, $c$ is the speed of light, and
$u$ is the relative velocity between $\sum$ and $\sum^{'}$ frame,
which move along $x$ and $x^{'}$ axes.
The velocity transformation is\\
\begin{eqnarray}
u_{x}&=&\frac{u_{x}^{'}+u
}{1+\frac{u u_{x}^{'}}{c_{m1}^{2}}}\nonumber\\
u_{y}&=&\frac{u_{y}^{'}\sqrt{1-\frac{u^{2}}{c_{m1}^{2}}}}
{1+\frac{u u_{x}^{'}}{c_{m}^{2}}}\nonumber\\
u_{z}&=&\frac{u_{z}^{'}\sqrt{1-\frac{u^{2}}{c_{m1}^{2}}}}
{1+\frac{u u_{x}^{'}}{c_{m1}^{2}}},
\end{eqnarray}
where $u_{x}$, $u_{y}$, $u_{z}$ and
$v=\sqrt{u_{x}^{2}+u_{y}^{2}+u_{z}^{2}}$ $(0\leq v<c_{m1})$ are a
particle velocity component and velocity amplitude in $\sum$
frame, $u_{x}^{'}$, $u_{y}^{'}$, $u_{z}^{'}$ and
$v'=\sqrt{{u'}_{x}^{2}+{u'}_{y}^{2}+{u'}_{z}^{2}}$ $(0\leq
v'<c_{m1})$ are the particle velocity component projection and
velocity amplitude in $\sum^{'}$ frame. Now, We can discuss the
problem of the speed of light. For two inertial reference frames
$\sum$ and $\sum^{'}$, the $\sum^{'}$ frame is a rest frame for
light, i.e., the two reference frames relative velocity $v$ is
equal to $c$. At the time $t=0$, a beam of light is emitted from
the origin $O$. When $u_{x}=c$, we obtain the result from Eq. (8),
\begin{equation}
u_{x}^{'}=0,
\end{equation}
when $u_{x}=-c$, we have
\begin{equation}
u_{x}^{'}=\frac{-c-c}{1+\frac{c^{2}}{c_{m1}^{2}}}=-2\frac{c
c_{m1}^{2}}{c^{2}+c_{m1}^{2}}>-2c.
\end{equation}
It shows that the invariant principle of light velocity is
violated for the second kind particle in the inertial reference of
light velocity movement, but the limit velocity $c_{m1}$ is
invariant. \\
For the second kind particle, the mass $m$, momentum $\vec{p}$ and
energy $E$ of a particle of rest mass $m_{0}$ and velocity
$\vec{v}$ are is
\begin{equation}
m=\frac{m_{0}}{\sqrt{1-\frac{\upsilon^{2}}{c_{m1}^{2}}}},
\end{equation}
\begin{equation}
\vec{p}=\frac{m_{0}}{\sqrt{1-\frac{\upsilon^{2}}{c_{m1}^{2}}}}\vec{v},
\end{equation}
and
\begin{equation}
E=mc_{m1}^{2},
\end{equation}
and the dispersion relation is
\begin{equation}
E^{2}=m_{0}^{2}c_{m1}^{4}+p^{2}c_{m1}^{2}.
\end{equation}
In the following, we shall study the nature of photon, and obtain
some new results. \\
From Eqs. (11)-(14), we obtain the photon mass, momentum and
energy at light velocity
\begin{equation}
m_{c}=\frac{m_{0}}{\sqrt{1-\frac{c^{2}}{c_{m1}^{2}}}},
\end{equation}
and
\begin{equation}
\vec{p}_{c}=\frac{m_{0}}{\sqrt{1-\frac{c^{2}}{c_{m1}^{2}}}}\vec{c},
\end{equation}
\begin{equation}
E=m_{v}c_{m1}^{2}=\frac{m_{0}}{\sqrt{1-\frac{v^{2}}{c_{m1}^{2}}}}c_{m1}^{2}=h\nu_{v},
\end{equation}
where $m_{v}$ is photon mass, $\nu_{v}$ is photon frequency at
velocity $v$ ($0\leq v<c_{m1}$). When $v=0$, we obtain the photon
rest energy $E_{0}$, it is
\begin{equation}
E_{0}=m_{0}c_{m1}^{2}=h\nu_{0},
\end{equation}
where $m_{0}$ is photon rest mass, $\nu_{0}$ is photon rest
frequency, we find photon has rest mass and rest frequency when
$v=c$, we obtain the photon energy at light velocity
\begin{equation}
E_{c}=m_{c}c_{m1}^{2}=\frac{m_{0}}
{\sqrt{1-\frac{c^{2}}{c_{m1}^2}}} c_{m1}^2=h\nu_{c},
\end{equation}
where $m_{c}$ and $\nu_{c}$ are photon mass and frequency at light
velocity $c$. From Eq. (19), we can obtain the photon rest mass
$m_{0}$ and light velocity mass $m_{c}$, they are
\begin{equation}
m_{0}=\frac{h\nu_{c}}{c_{m1}^2}\sqrt{1-\frac{c^2}{c_{m1}^2}},
\end{equation}
and
\begin{equation}
m_{c}=\frac{h\nu_{c}}{c_{m1}^2},
\end{equation}
From Eqs. (17) and (18), we have
\begin{equation}
\nu_{v}=\frac{m_{0}c_{m1}^{2}}{h\sqrt{1-\frac{v^{2}}{c_{m1}^{2}}}}=\frac{\nu_{0}}{\sqrt{1-\frac{v^{2}}{c_{m1}^{2}}}}.
\end{equation}
The Eq. (22) gives the relation of a photon's frequency with its
velocity, and we find photon movement frequency $\nu_{v}$ is
larger than its rest frequency $\nu_{0}$.

When $v=c$, we obtain the photon frequency $\nu_{c}$ at light
velocity $c$
\begin{equation}
\nu_{c}=\frac{\nu_{0}}{\sqrt{1-\frac{c^{2}}{c_{m1}^{2}}}}.
\end{equation}
From Eq. (22) and (23), we obtain the ratio between the photon
frequency at arbitrary velocity $v$ and the frequency at light
velocity $c$
\begin{equation}
\frac{\nu_{v}}{\nu_{c}}=\frac{\sqrt{c_{m1}^2-c^2}}{\sqrt{c_{m1}^2-v^2}}.
\end{equation}
Substituting Eq. (24) into the relation $\nu_{v}\lambda=v$, i.e.,
\begin{equation}
\frac{m_{0}c_{m1}^{2}}{h}\frac{1}{\sqrt{1-\frac{v^{2}}{c_{m1}^{2}}}}\lambda=v,
\end{equation}
we obtain
\begin{equation}
\lambda=\frac{hv}{m_{0}c_{m1}^{2}}\sqrt{1-\frac{v^{2}}{c_{m1}^{2}}},
\end{equation}
The Eq. (26) gives the relation between photon wavelength and its
velocity $v$. When $v=0$, the light wavelength $\lambda=0$, i.e.,
the photon hasn't wavelength when it rests, but it has rest mass
$m_{0}$ and rest frequency $\nu_{0}$.\\
From Eq. (17) and (19), we have
\begin{equation}
c_{m1}= \sqrt{\frac{h\nu_{v}}{m_{v}}},
\end{equation}
and
\begin{equation}
c_{m1}= \sqrt{\frac{h\nu_{c}}{m_{c}}}
\end{equation}
From Eqs. (27) and (28), we can calculate the limit velocity
$c_{m1}$ if we can measure the photon mass $m_{v}$($m_{c}$) when
its frequency is $\nu_{v}$($\nu_{c}$) at arbitrary velocity $v$
(light velocity $c$). All photon possess a finite mass and their
physical implications have been discussed by many theories and
experiments [20, 21, 22]. In Ref. [21, 22], the experiment were
made by laser, and it determined the range of the photon mass is
$10^{-6}eV<m_{\nu}<10^{-4}eV$. We know the laser frequency is in
the range of $8.9\times 10^{13}Hz$ $\sim$ $9.23\times 10^{14}Hz$.
From Eq. (28), we can estimate the limit velocity $c_{m1}$. It is
in the rang of:
\begin{equation}
\sqrt{\frac{6.626\times10^{-34}\times
8.9\times10^{13}}{10^{-4}\times1.6\times10^{-19}}}c\leq c_{m1}=
\sqrt{\frac{h\nu_{c}}{m_{c}}}\leq\sqrt{\frac{6.626\times10^{-34}\times
9.23\times10^{14}}{10^{-6}\times1.6\times10^{-19}}}c,
\end{equation}
i.e.,
\begin{equation}
60c\leq c_{m1}\leq 2000c.
\end{equation}
If we take the middle experiment value, i.e., $m_{c}=10^{-5}$eV,
$\nu_{c}=5.5\times 10^{14}$Hz, the limit velocity is
\begin{equation}
c_{m1}=\sqrt{\frac{h\nu_{c}}{m_{{c}}}}= \sqrt{\frac{6.626\times
10^{-34}\times5.5\times10^{14}}
{10^{-5}\times1.6\times10^{-19}}}=480c.
\end{equation}
In experiment [23], E. Fomalont etal. have found light of
frequency $\nu=43GeV$ passing near the solar limb, the photon mass
upper limit is $3.5\times10^{-11}MeV$. We can estimate the lower
limit of limit velocity $c_{m}$, it is
\begin{equation}
c_{m1}=\sqrt{\frac{h\nu_{v}}{m_{{v}}}}>\sqrt{\frac{6.626\times
10^{-34}\times43\times10^{9}}
{3.5\times10^{-11}\times10^{6}\times1.6\times10^{-19}}}=22c.
\end{equation}
In Ref. [24], the experiment measured the superluminal velocity is
$310c$. In Refs. [25-26], the experiments measured signal velocity
were $4.7c$ for the microwave and $1.7c$ for single photon.

\section * {3. The space-time transformation and mass-velocity relation
for the third kind of particle ($c\leq \upsilon < c_{m2}$)}

In the following, we will give the space-time relation in two
inertial reference frames $\sum$ and $\sum^{'}$ for the third kind
of particle, which its velocity is in the range of $c\leq \upsilon
< c_{m2}$. We think the muonic neutrinos and tachyon travels
faster than light, but its velocity can not be infinity. So, we
can assume there is a limit velocity $c_{m2}$ for the third kind
of particle, and we also give two postulates:

1. The Principle of Relativity: All laws of nature are the same in
all inertial reference frames.

2. The Invariant Principle of Limit Velocity: There is a limit
velocity $c_{m2}$ in nature, and the $c_{m2}$ is invariant in all
inertial reference frames.

From the two postulates, we can obtain the new space-time
transformation and velocity transformation for the third kind of
particle. When we replace $c$ with $c_{m2}$, we can obtain the new
transformation relation from the Lorentz transformation Eq. (1),
they are
\begin{eqnarray}
x&=&\frac{x^{'}+u
t^{'}}{\sqrt{1-\frac{u^{2}}{c_{m2}^{2}}}}\nonumber\\
 y&=&y^{'}\nonumber\\
z&=&z^{'}\nonumber\\
t&=&\frac{t^{'}+\frac{u}{c_{m2}^{2}}x^{'}}{\sqrt{1-\frac{u^{2}}{c_{m2}^{2}}}},
\end{eqnarray}
where $x$, $y$, $z$, $t$ are space-time coordinates in $\sum$
frame, $x^{'}$, $y^{'}$, $z^{'}$, $t^{'}$ are space-time
coordinates in $\sum^{'}$ frame, $c_{m2}$ is the limit velocity,
and $u$ is the relative velocity between $\sum$ and $\sum^{'}$
frame, which move along $x$ and $x^{'}$ axes.
The velocity transformation is\\
\begin{eqnarray}
u_{x}&=&\frac{u_{x}^{'}+u
}{1+\frac{u u_{x}^{'}}{c_{m}^{2}}}\nonumber\\
u_{y}&=&\frac{u_{y}^{'}\sqrt{1-\frac{u^{2}}{c_{m2}^{2}}}}
{1+\frac{u u_{x}^{'}}{c_{m2}^{2}}}\nonumber\\
u_{z}&=&\frac{u_{z}^{'}\sqrt{1-\frac{u^{2}}{c_{m2}^{2}}}}
{1+\frac{u u_{x}^{'}}{c_{m2}^{2}}},
\end{eqnarray}
where $u_{x}$, $u_{y}$, $u_{z}$ and
$v=\sqrt{u_{x}^{2}+u_{y}^{2}+u_{z}^{2}}$ $(c\leq v<c_{m2})$ are a
particle velocity component and velocity in $\sum$ frame,
$u_{x}^{'}$, $u_{y}^{'}$, $u_{z}^{'}$ and
$v'=\sqrt{{u'}_{x}^{2}+{u'}_{y}^{2}+{u'}_{z}^{2}}$ $(c\leq
v'<c_{m2})$ are the particle velocity component and velocity in
$\sum^{'}$ frame.

In the following, we will give the new relation of particle mass
$m$ with its velocity $v$. We can consider the collision between
two identical particle. It is shown in Figure 1.
\begin{center}
\setlength{\unitlength}{0.1in}
\begin{figure}
\begin{picture}(100,10)
\put(20,4){\vector(1,0){20}}
  \put(38,2){\makebox(2,1)[l]{$x$}}
  \put(40,2.3){\makebox(2,1)[l]{$x^{\prime}$}}
\put(23,9){\vector(1,0){4}}
  \put(24,10){\makebox(2,1)[c]{$\vec{v}$}}
\put(22,6){\vector(1,0){2}}
  \put(22,6.7){\makebox(2,1)[c]{$\vec{v_{1}}$}}
  \put(21,4.5){\makebox(2,1)[c]{$m_{1}$}}
\put(26,6){\vector(1,0){2}}
  \put(26,6.7){\makebox(2,1)[c]{$\vec{v_{2}}$}}
  \put(25,4.5){\makebox(2,1)[c]{$m_{2}$}}
\put(20,4){\vector(0,1){10}}
  \put(18,13){\makebox(2,1)[c]{$y$}}
\put(30,4){\vector(0,1){10}}
  \put(28,13){\makebox(2,1)[c]{$y^{\prime}$}}
\put(30,4){\vector(-1,-1){5}}
  \put(24,-2.5){\makebox(2,1)[c]{$z^{\prime}$}}
\put(20,4){\vector(-1,-1){5}}
  \put(14,-2.5){\makebox(2,1)[c]{$z$}}
  \put(30,2.6){\makebox(2,1)[l]{$o^{\prime}$}}
  \put(20,2.6){\makebox(2,1)[l]{$o$}}
  \put(32,12){\makebox(2,1)[c]{$\Sigma^{\prime}$}}
  \put(22,12){\makebox(2,1)[c]{$\Sigma$}}
 \put(22,6){\circle*{0.5}}
 \put(26,6){\circle*{0.5}}
\end{picture}

\caption{The $\Sigma$ is the laboratory system, $\Sigma^{\prime}$
is the mass-center system for two particles $m_1$ and $m_2$, and
the two inertial frames relative velocity is $v$.} \label{moment}
\end{figure}
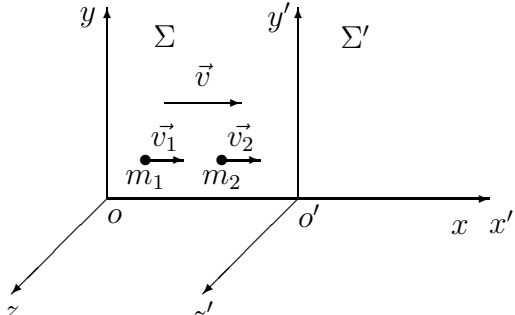
\end{center}
\newpage
The $\Sigma$ is the laboratory system, and $\Sigma^{\prime}$ is
the mass-center system of two particles $m_{1}$ and $m_{2}$. In
$\Sigma$ system, the velocity of two particles $m_{1}$ and $m_{2}$
are $\vec{v_{1}}$ and $\vec{v_{2}}$ $(c\leq v_{2}< v_{1}<c_{m2})$,
which are along with $x (x^{\prime})$ axis, and they are
$v^{\prime}$ and $-v^{\prime}$ in $\Sigma^{\prime}$ system. After
collision, the two particles velocities are all $v$ $(c\leq
v<c_{m2}$ in $\Sigma$ system. The momentum was conserved in this
process:
\begin{equation}
m_{1}v_{1}+m_{2}v_{2}=(m_{1}+m_{2})v.
\end{equation}
According to equation (34), there is
\begin{eqnarray}
v_{1}=\frac{v^{\prime}+v}{1+\frac{v v^{\prime}}{c_{m2}^{2}}} \nonumber\\
v_{2}=\frac{-v^{\prime}+v}{1-\frac{v v^{\prime}}{c_{m2}^{2}}}.
\end{eqnarray}
From equations (35) and (36), we get
\begin{equation}
m_{1}(1-\frac{v v^{\prime}}{c_{m2}^{2}})=m_{2}(1+\frac{v
v^{\prime}}{c_{m2}^{2}}).
\end{equation}
From equation (34), we can obtain
\begin{equation}
1-\frac{u_{x}^{\prime}u}{c_{m2}^{2}}=\frac{\sqrt{1-\frac{{v}^{
2}}{c_{m2}^{2}}} \sqrt{1-\frac{u^{2}}{c_{m2}^{2}}}}
{\sqrt{1-\frac{{v'}^{2}}{c_{m2}^{2}}}},
\end{equation}
where $v=\sqrt{u_{x}^{2}+u_{y}^{2}+u_{z}^{2}}$ and
$v'=\sqrt{{u'}_{x}^{2}+{u'}_{y}^{2}+{u'}_{z}^{2}}$. For the
particle $m_{1}$, the equation (38) becomes
\begin{equation}
1+\frac{v^{\prime}v}{c_{m2}^{2}}=\frac{\sqrt{1-\frac{v^{\prime2}}{c_{m2}^{2}}}
\sqrt{1-\frac{v^{2}}{c_{m2}^{2}}}}
{\sqrt{1-\frac{v_{1}^{2}}{c_{m2}^{2}}}}.
\end{equation}
For the particle $m_{2}$, the equation (38) becomes
\begin{equation}
1-\frac{v^{\prime}v}{c_{m2}^{2}}=\frac{\sqrt{1-\frac{v^{\prime2}}{c_{m2}^{2}}}
\sqrt{1-\frac{v^{2}}{c_{m2}^{2}}}}
{\sqrt{1-\frac{v_{2}^{2}}{c_{m2}^{2}}}}.
\end{equation}
By substituting equations (39) and (40) into (37), we get
\begin{equation}
m(v_{1})\sqrt{1-\frac{v_{1}^{2}}{c_{m2}^{2}}}=
m(v_{2})\sqrt{1-\frac{v_{2}^{2}}{c_{m2}^{2}}}=
m(c)\sqrt{1-\frac{c^{2}}{c_{m2}^{2}}}=constant,
\end{equation}
where $m(c)$ is the particle mass when its velocity is light
velocity $c$.\\
For arbitrary velocity $v$ ($c\leq v<c_{m2}$), we have
\begin{equation}
m(v)\sqrt{1-\frac{v^{2}}{c_{m2}^{2}}}=m(c)\sqrt{1-\frac{c^{2}}{c_{m2}^{2}}},
\end{equation}
and hence
\begin{equation}
m(v)=m_{c}\sqrt{\frac{c_{m2}^{2}-c^{2}}{c_{m2}^{2}-v^{2}}},
\end{equation}
where $m_{c}=m(c)$. The equation (43) is the relation between
tachyon mass $m$ and its velocity $v$ ($c\leq v<c_m$).

From Eqs. (11)-(14), we can obtain the third kind particle energy,
momentum and dispersion relation by replacing $c$ with $c_{m2}$,
and replacing $m_{0}$ with $m_{c}$, they are
\begin{equation}
E=\frac{m_{c}c_{m2}^{2}}{\sqrt{1-\frac{v^{2}}{c_{m2}^{2}}}},
\end{equation}
\begin{equation}
\vec{p}=\frac{m_{c}\vec{v}}{\sqrt{1-\frac{v^{2}}{c_{m2}^{2}}}},
\end{equation}
and
\begin{equation}
E^{2}-p^{2}c_{m2}^{2}=m_{c}^{2}c_{m2}^{4}.
\end{equation}
By substituting Eq. (43) into (44), we can obtain the relation of
mass-energy.
\begin{equation}
E=\frac{m(v)}{\sqrt{c_{m2}^{2}-c^{2}}}c_{m2}^{3}.
\end{equation}
From Eqs. (43) and (47), we can analysis the OPERA muonic neutrino
experiment. and estimated the velocity limit $c_{m2}$ for the
third kind particle. Otherwise, we can calculate the muonic
neutrino mass.

\section * {4. The Lorentz group and extended Lorentz group}

\hspace{0.3in} For the first kind particle, we can obtain the
invariant interval $ds$ by the invariant principle of light
velocity, it is
\begin{equation}
ds^{2}=c^{2}dt^{2}-dx^{2}-dy^{2}-dz^{2},
\end{equation}
\begin{equation}
ds'^{2}=c^{2}dt'^{2}-dx'^{2}-dy'^{2}-dz'^{2},
\end{equation}
and
\begin{equation}
ds^{2}=ds'^{2}.
\end{equation}
The Lorentz transformation is a line transformation in 4-dimension
space-time, which satisfies the invariant of internal, it is
\begin{equation}
 x'_{\mu}=a_{\mu\nu}x_{\nu}
\end{equation}
where $\mu,\nu$=0.1.2.3, and $x_{0}$=$ct$, $x_{1}$=$x$,
$x_{2}$=$y$, $x_{3}$=$z$. \\
The matrix form of Eq. (51) is
\begin{eqnarray}
\left ( \begin{array}{llll}
  x'_{0} \\
  x'_{1}\\
  x'_{2}\\
  x'_{3}\\
\end{array}
   \right )=\left ( \begin{array}{llll}
a_{00}\hspace{0.3in}a_{01}\hspace{0.3in}a_{02}\hspace{0.3in}a_{03}\\
a_{10}\hspace{0.3in}a_{11}\hspace{0.3in}a_{12}\hspace{0.3in}a_{13}\\
a_{20}\hspace{0.3in}a_{21}\hspace{0.3in}a_{22}\hspace{0.3in}a_{23}\\
a_{30}\hspace{0.3in}a_{31}\hspace{0.3in}a_{32}\hspace{0.3in}a_{33}\\
\end{array}
   \right )\left ( \begin{array}{llll}
  x_{0} \\
  x_{1}\\
  x_{2}\\
  x_{3}\\
\end{array}
   \right ),
\end{eqnarray}
and Lorentz transformation $a_{\mu\nu}$ satisfies  the orthogonal
relation
\begin{equation}
a_{\mu\nu}a_{\mu\rho}=\delta_{\upsilon\rho},
\end{equation}
the aggregate of orthogonal transformation $a_{\mu\nu}$
constitutes a group, which is Lorentz group. \\
For $v$ in the x-direction, the special Lorentz transformation is
\begin{eqnarray}
a=\left ( \begin{array}{llll}
   \frac{1}{\sqrt{1-\frac{v^{2}}{c^{2}}}}\hspace{0.45in} 0\hspace{0.5in}0\hspace{0.5in}
   i\frac{v}{c}\frac{1}{\sqrt{1-\frac{v^{2}}{c^{2}}}}\\
               \hspace{0.3in}0\hspace{0.5in}1\hspace{0.5in} 0\hspace{0.75in} 0\\
               \hspace{0.3in}0\hspace{0.5in} 0\hspace{0.5in} 1\hspace{0.75in} 0\\
 -i\frac{v}{c}\frac{1}{\sqrt{1-\frac{v^{2}}{c^{2}}}}\hspace{0.20in}0\hspace{0.5in}0\hspace{0.5in}
 \frac{1}{\sqrt{1-\frac{v^{2}}{c^{2}}}}\\
\end{array}
   \right ).
\end{eqnarray}
For the second and third kind particles, we can also obtain the
invariant interval $ds$ by the invariant principle of limit
velocity, they are
\begin{equation}
ds^{2}=c_{mi}^{2}dt^{2}-dx^{2}-dy^{2}-dz^{2},
\end{equation}
\begin{equation}
ds'^{2}=c_{mi}^{2}dt'^{2}-dx'^{2}-dy'^{2}-dz'^{2},
\end{equation}
and
\begin{equation}
ds^{2}=ds'^{2}.
\end{equation}
Where $c_{mi} (i=1,2)$ is the limit velocity, and $c_{m1}$,
$c_{m2}$ are the limit velocity of second and third kind
particles. The transformation is
\begin{equation}
x'_{\mu}=b_{\mu\nu}x_{\nu},
\end{equation}
the matrix form of Eq. (58) is
\begin{eqnarray} \left (
\begin{array}{llll}
  x'_{0} \\
  x'_{1}\\
  x'_{2}\\
  x'_{3}\\
\end{array}
   \right )=\left ( \begin{array}{llll}
b_{00}\hspace{0.3in}b_{01}\hspace{0.3in}b_{02}\hspace{0.3in}b_{03}\\
b_{10}\hspace{0.3in}b_{11}\hspace{0.3in}b_{12}\hspace{0.3in}b_{13}\\
b_{20}\hspace{0.3in}b_{21}\hspace{0.3in}b_{22}\hspace{0.3in}b_{23}\\
b_{30}\hspace{0.3in}b_{31}\hspace{0.3in}b_{32}\hspace{0.3in}b_{33}\\
\end{array}
   \right )\left ( \begin{array}{llll}
  x_{0} \\
  x_{1}\\
  x_{2}\\
  x_{3}\\
\end{array}
   \right ),
\end{eqnarray}
and the transformation $b_{\mu\nu}$ satisfies the orthogonal
relation
\begin{equation}
b_{\mu\nu}b_{\mu\rho}=\delta_{\upsilon\rho}.
\end{equation}
The aggregate of orthogonal transformation $b_{\mu\nu}$ constitute
a group, which is extended Lorentz group.\\
For $v$ in the x-direction, the special extended Lorentz
transformation is
\begin{eqnarray}
b=\left ( \begin{array}{llll}
   \hspace{0.2in}\frac{1}{\sqrt{1-\frac{v^{2}}{c_{mi}^{2}}}}\hspace{0.62in}
   0\hspace{0.5in}0\hspace{0.5in}i\frac{v}{c_{mi}}\frac{1}{\sqrt{1-\frac{v^{2}}{c_{mi}^{2}}}}\\
   \hspace{0.5in}0\hspace{0.75in}1\hspace{0.5in} 0\hspace{0.75in} 0\\
   \hspace{0.5in}0\hspace{0.75in} 0\hspace{0.5in} 1\hspace{0.75in} 0\\
 -i\frac{v}{c_{mi}}\frac{1}{\sqrt{1-\frac{v^{2}}{c_{mi}^{2}}}}\hspace{0.46in}0\hspace{0.5in}0
 \hspace{0.55in}\frac{1}{\sqrt{1-\frac{v^{1}}{c_{mi}^{2}}}}\\
\end{array}
   \right ).
\end{eqnarray}
\section * {5. The relativistic dynamics for the second and third kinds particles}

For the first kind particle ($0\leq v<c$), we know the 4-force is
defined as:
\begin{equation}
K_{\mu}=\frac{dp_{\mu}}{d\tau}=(\vec{K}, iK_{4}),
\end{equation}
the "ordinary" force $\vec{K}$ is
\begin{equation}
\vec{K}=\frac{d\vec{p}}{dt}
\frac{dt}{d\tau}=\frac{1}{\sqrt{1-\frac{v^{2}}{c^{2}}}}\frac{d\vec{p}}{dt},
\end{equation}
while the fourth component
\begin{eqnarray}
K_{4}&=&\frac{dp_{4}}{d\tau}=\frac{1}{c}\frac{dE}{d\tau} \nonumber\\
&=&\frac{1}{c}\vec{v}\cdot \vec{K},
\end{eqnarray}
so
\begin{equation}
K_{\mu}=(\vec{K}, \frac{i}{c}\vec{v}\cdot \vec{K}),
\end{equation}
the covariant equation for a particle are
\begin{equation}
\vec{K}=\frac{d\vec{p}}{d\tau},
\end{equation}
\begin{equation}
\vec{K}\cdot \vec{v}=\frac{dE}{d\tau}.
\end{equation}
\begin{equation}
\sqrt{1-\frac{v^{2}}{c^{2}}}\vec{K}\cdot \vec{v}=\frac{dE}{dt},
\end{equation}
we define force $\vec{F}$ as
\begin{equation}
\vec{F}=\sqrt{1-\frac{v^{2}}{c^{2}}}\vec{K}.
\end{equation}
From Eqs. (66)-(68), we have the relativistic dynamics equations
for a particle are
\begin{equation}
\vec{F}=\frac{d\vec{p}}{dt},
\end{equation}
\begin{equation}
\vec{K}\cdot \vec{v}=\frac{dE}{dt}.
\end{equation}

For the second kind particle ($0\leq v<c_{m1}$), the relativistic
dynamics equations are Eqs. (70) and (71), but some physical
quantities should be modified \\
The 4-momentum and 4-force are
\begin{equation}
p_{\mu}=m_{0}U_{\mu}=(\vec{p}, \frac{i}{c_{m1}}E),
\end{equation}
and
\begin{equation}
K_{\mu}=(\vec{K}, iK_{4}),
\end{equation}
with
$\vec{p}=\frac{m_{0}\vec{v}}{\sqrt{1-\frac{v^{2}}{c_{m1}^{2}}}}$,
$p_{4}=\frac{m_{0}c_{m1}}{\sqrt{1-\frac{v^{2}}{c_{m1}^{2}}}}$,
$\vec{K}=\frac{d\vec{p}}{dt}\frac{1}{\sqrt{1-\frac{v^{2}}{c_{m1}^{2}}}}$
and $K_{4}=\frac{1}{c_{m1}}\vec{v}\cdot \vec{K}$.\\

For the third kind particle ($c\leq v<c_{m2}$), the relativistic
dynamics equations are also Eqs. (70) and (71), and some physical
quantities also should be modified \\
The 4-momentum and 4-force are
\begin{equation}
p_{\mu}=m_{c}U_{\mu}=(\vec{p}, \frac{i}{c_{m2}}),
\end{equation}
and
\begin{equation}
K_{\mu}=(\vec{K}, iK_{4}),
\end{equation}
with
$\vec{p}=\frac{m_{c}\vec{v}}{\sqrt{1-\frac{v^{2}}{c_{m2}^{2}}}}$,
$p_{4}=\frac{m_{c}c_{m2}}{\sqrt{1-\frac{v^{2}}{c_{m2}^{2}}}}$,
$\vec{K}=\frac{d\vec{p}}{dt}\frac{1}{\sqrt{1-\frac{v^{2}}{c_{m2}^{2}}}}$
and $K_{4}=\frac{1}{c_{m2}}\vec{v}\cdot \vec{K}$.

\section * {6. The quantum wave equation for the second and third kinds particles}

For the first kind particle ($0\leq v<c$), we express $E$ and
$\vec{p}$ as operators:
\begin{eqnarray}
E\rightarrow i\hbar\frac{\partial}{\partial t} \nonumber\\
\vec{p}\rightarrow -i\hbar \nabla,
\end{eqnarray}
we can obtain the quantum wave equation of spin $0$ particle from
Eq. (6)
\begin{equation}
[\frac{\partial^{2}}{\partial
t^{2}}-c^{2}\nabla^{2}+\frac{m_{0}^{2}c^{4}}{\hbar^{2}}]\Psi(\vec{r},
t)=0,
\end{equation}
and we can obtain the quantum wave equation of spin $\frac{1}{2}$
particle
\begin{equation}
i\hbar \frac{\partial}{\partial t}\Psi=[-i\hbar c
\vec{\alpha}\cdot \vec{\nabla}+m_{0}c^{2}\beta]\Psi,
\end{equation}
where $\vec{\alpha}$ and $\beta$ are matrixes
\[
\vec{\alpha}= \left (
\begin {array} {cc}
0 & \vec{\sigma} \\
\vec{\sigma} & 0
\end{array} \right ),
\]
and
\[
\beta= \left (
\begin {array} {cc}
I & 0 \\
0 & -I
\end{array} \right ),
\]
where $\vec{\sigma}$ are Pauli matrixes, and $I$ is unit matrix of
$2\times 2$.

For the second kind particle ($0\leq v<c_{m1}$), we can obtain the
quantum wave equation of spin $0$ particle from Eq. (14)
\begin{equation}
[\frac{\partial^{2}}{\partial
t^{2}}-c_{m1}^{2}\nabla^{2}+\frac{m_{0}^{2}c_{m1}^{4}}{\hbar^{2}}]\Psi(\vec{r},
t)=0,
\end{equation}
and we can obtain the quantum wave equation of spin $\frac{1}{2}$
particle
\begin{equation}
i\hbar \frac{\partial}{\partial t}\Psi=[-i\hbar c_{m1}
\vec{\alpha}\cdot \vec{\nabla}+m_{0}c_{m1}^{2}\beta]\Psi.
\end{equation}
For the third kind particle ($c\leq v<c_{m2}$), we can obtain the
quantum wave equation of spin $0$ particle from Eq. (46)
\begin{equation}
[\frac{\partial^{2}}{\partial
t^{2}}-c_{m2}^{2}\nabla^{2}+\frac{m_{c}^{2}c_{m2}^{4}}{\hbar^{2}}]\Psi(\vec{r},
t)=0,
\end{equation}
and we can obtain the quantum wave equation of spin $\frac{1}{2}$
particle
\begin{equation}
i\hbar \frac{\partial}{\partial t}\Psi=[-i\hbar c_{m2}
\vec{\alpha}\cdot \vec{\nabla}+m_{c}c_{m2}^{2}\beta]\Psi.
\end{equation}

\section * { 7. Numerical analysis for OPERA experiment}\vskip 8pt

At OPERA muonic neutrino experiment, when the average neutrino
energy $17$ GeV, the relative difference of the velocity of the
muon neutrinos $v$ is:
$(v-c)/c=(2.48\pm0.28(stat)\pm0.30(sys))\times 10^{-5}$, and the
neutrinos produced at CERN are tachyons with mass $m$, after
having travelled a distance $L\approx730$ km, their associated
early arrival time is $\delta t = \frac{L}{c} \frac{v-c}{c}$, with
$\frac{L}{c}\approx2.4$ ms. Consider two tachyonic neutrino beams
of energy $E_{1}$ and $E_{2}$, with $E_{1}<E_{2}$ for
definiteness. The OPERA Collaborations considers two sample
neutrino beams with mean energy equal to $E_{1} = 13.9$ GeV and
$E_{2} = 42.9$ GeV respectively. The experimental values of the
associated early arrival times are respectively $\delta t_{1} =
(53.1 \pm 18.8(stat) \pm 7.4(sys))$ ns and $\delta t_{2} = (67.1
\pm 18.2(stat)\pm7.4(sys))$ ns. The OPERA experiment data can be
concluded as follows:
\begin{equation}
E_{1}=13.9 GeV,\hspace{0.15in} v_{1 min}=(3 \times
10^{8}+3362)m/s, \hspace{0.15in} v_{1 max}=(3\times 10^{8}+9913)
m/s,
\end{equation}
\begin{equation}
E_{2}=17 GeV,\hspace{0.15in} v_{2 min}=(3\times 10^{8}+5700)m/s,
\hspace{0.15in} v_{2 max}=(3\times 10^{8}+9180)m/s,
\end{equation}
\begin{equation}
E_{3}=42.9 GeV,\hspace{0.15in} v_{3 min}=(3\times 10^{8}+5188)m/s,
\hspace{0.15in} v_{3 max}=(3\times 10^{8}+11588)m/s,
\end{equation}

using the experiment data above, we can calculate the limit
velocity $c_{m2}$ and muonic neutrinos mass by Eqs.(43)-(47). \\
With Eq. (44), we have
\begin{equation}
\frac{E_{i}}{E_{j}}=\frac{\sqrt{c_{m2}^{2}-v_{j}^{2}}}{\sqrt{c_{m2}^{2}-v_{i}^{2}}},
\hspace{0.15in} (i, j=1, 2, 3)
\end{equation}

Table 1: Different energy muonic neutrino according to its
velocity, mass and the limit velocity.
\begin{center}
\begin{tabular}{|c|c|c|c|c|}\hline
E (GeV)& $v_{min}$(m/s) & $v_{max}$(m/s) & $m_{v}$ $(GeV
s^{2}/m^{2} )$& $m_{c}$ $(GeV s^{2}/m^{2} )$
\\
\hline 13.9 & $3\times 10^{8}+3363$ & $3\times 10^{8}+9912$ &
$(0.48\sim 1.82)(\times 10^{-18})$ & $(0.772\sim 1.671)(\times
10^{-18})$  \\
\cline{1-5} 17 & $3\times 10^{8}+5700$ & $3\times 10^{8}+9180$ &
$(1.03\sim 2.23)(\times 10^{-18})$ & $(0.705\sim 1.527)(\times
10^{-18})$ \\
\cline{1-5} 42.9 & $3\times 10^{8}+5188$ & $3\times 10^{8}+11588$
& $(2.60\sim 5.63)(\times 10^{-18})$ & $(1.161 \sim  2.511)(\times
10^{-18})$   \\
\hline
\end{tabular}
\end{center}
Substituting Eqs.(83)-(85) into (86), we can calculate the limit
velocity $c_{m2}$, it is in the range of $(3\times
10^{8}+15534)^{+5390}_{-11091}$m/s. From Eq. (47), we can
calculate the muonic neutrinos mass $m(v)$ at different energy.
Otherwise, we can calculate the muonic neutrinos mass $m(c)$ at
light velocity by Eq. (43). All the calculation results are shown
in Table 1, where the first column is muonic neutrinos energy, the
second and third column are corresponding to different energy
neutrinos minimum and maximum velocity, the forth column is
corresponding to different energy neutrinos mass, the final column
is muonic neutrinos mass at light velocity $c$. From Table 1, we
can find the muonic neutrinos mass $m(c)$ is in the range of
$(1.161\sim1.527)\times10^{-18}$$GeV s^{2}/m^{2}$, and the muonic
neutrinos mass increase when its energy increase.

\section * {8. Conclusion}

In this paper, an extended relativity theory is proposed. We think
all particles can be divided into three kinds: The first kind of
particle is its velocity in the range of $0\leq v < c$, e.g.
electron, atom, molecule and so on. The second kind of particle is
its velocity in the range of $0\leq v < c_{m1}$, e.g. photon. The
third kind of particle is its velocity in the range of $c\leq v <
c_{m2}$, e.g. tachyon, and muonic neutrinos. The first kind of
particle is described by the special relativity. With the extended
relativity theory, we can describe the second and third kinds
particles, and obtain some new results. Otherwise, we analysis the
OPERA experiment data and calculate the muonic neutrinos mass.


\newpage
\begin{thebibliography}{99}

\bibitem{s1}
Opera Collaboration, Measurement of the neutrino velocity with the
OPERA detector in the CNGS beam, arXiv:1109.4897.

\bibitem{s2}
P. Adamson et al. [ MINOS Collaboration ], Measurement of neutrino
velocity with the MINOS detectors and NuMI neutrino beam, Phys.
Rev. D76 (2007) 072005.

\bibitem{s3}
G. R. Kalb eisch, N. Baggett, E. C. Fowler, J. Alspector,
Experimental Comparison Of Neu- trino, Anti-neutrino, And Muon
Velocities, Phys. Rev. Lett. 43 (1979) 1361.

\bibitem{s4}
J. R. Ellis, N. E. Mavromatos, D. V. Nanopoulos, A. S. Sakharov,
Int. J. Mod. Phys. A19, 4413-4430 (2004).

\bibitem{s5}
V. A. Kostelecky, S. Samuel, Phys. Rev.D39, 683 (1989).

\bibitem{s6}
D. J. H. Chung, E. W. Kolb, A. Riotto, Phys. Rev. D65, 083516
(2002).

\bibitem{s7}
C. Csaki, J. Erlich, C. Grojean, Nucl. Phys. B604, 312- 342
(2001).

\bibitem{s8}
P. Horava, Phys. Rev. D79, 084008 (2009).

\bibitem{s9}
G. F. Giudice, M. Raidal, A. Strumia, Phys. Lett. B690, 272-279
(2010).

\bibitem{s10}
R.Cameron et al, Phys. Rev. D {\bf 47} (1993) 3707.

\bibitem{s11}
H. Gies, J. Jaeckel and A. Ringwald, Europhys. Lett. {\bf 76}
(2006) 794.

\bibitem{s12}
M. Ahlers, H. Gies, J. Jaeckel and A. Ringwald, Phys. Rev. D {\bf
75} (2007) 035011.

\bibitem{s13}
M. S. Bigelow, N. N. Lepeshkin and W. R. Boyd, Science.  {\bf301}
(2003) 200

\bibitem{s14}
L. J. Wang, A. Kuzmich and A. Dogariu, Nature.  {\bf406} (2000)
277

\bibitem{s15}
M. W. Mitchell and R. Y. Chiao, Am. J. Phys.  {\bf66} (1998) 14

\bibitem{s16}
M. A. Steinberg, P. G. Kwiat and R. Y. Chiao, Phys. Rev. Lett.
{\bf 71} (1993) 708

\bibitem{s17}
I. Alexeev, K. Y. Kim and H. M. Milchbery, Phys. Rev. Lett. {\bf
88} (2002) 073901

\bibitem{s18}
O.M.P. Bilaniuk, V.K. Deshpand and E.S.G. Sudarshan, Am. J. Phys.
{\bf 30}(no. 10) (1962) 718.

\bibitem{s19}
O.M.P. Bilaniuk and E.S.G. Sudarshan, Physics Today {\bf 22}(no.
10) (1969) 843.

\bibitem{s20}
R.Cameron et al, Phys. Rev. D {\bf 47} (1993) 3707.

\bibitem{s21}
H. Gies, J. Jaeckel and A. Ringwald, Europhys. Lett. {\bf 76}
(2006) 794.

\bibitem{s22}
M. Ahlers, H. Gies, J. Jaeckel and A. Ringwald, Phys. Rev. D {\bf
75} (2007) 035011.

\bibitem{s23}
E. Fomalont et al., Astrophys. J. 699, (2009) 1395.

\bibitem{s24}
L. J. Wang, A. Kuzmich and A. Dogariu, Nature {\bf 406} (2000)
277.

\bibitem{s25}
A. M. Steinberg, P. G. Kwiat and R. Y. Chiao, Phys. Rev. Lett.
{\bf 71} (1993) 708.

\bibitem{s26} R. Y. Chiao, P. G. Kwiat and A. M. Steinberg, Sci Am.
{\bf 269} (1993) 52.


\end{thebibliography}
\end{document}